\documentclass[journal]{IEEEtran}
\IEEEoverridecommandlockouts

\usepackage{cite}
\usepackage{amsmath,amssymb,amsfonts}
\usepackage{algorithmic}
\usepackage{graphicx}
\usepackage{textcomp}
\usepackage{xcolor}
\def\BibTeX{{\rm B\kern-.05em{\sc i\kern-.025em b}\kern-.08em
    T\kern-.1667em\lower.7ex\hbox{E}\kern-.125emX}}

\usepackage{algorithm}
\usepackage{array}
\usepackage{graphicx}
\usepackage{url}
\usepackage{verbatim}
\usepackage{cite}
\usepackage{stfloats}
\usepackage{caption}
\usepackage{subcaption}
\usepackage{multirow}
\usepackage{multicol}
\usepackage{comment}
\usepackage{upgreek}
\usepackage{tikz}
\renewcommand\fbox{\fcolorbox{red}{white}}
\newcommand\submittedtext{%
  \footnotesize This work has been submitted to the IEEE for possible publication. Copyright may be transferred without notice, after which this version may no longer be accessible.}

\newcommand\submittednotice{%
\begin{tikzpicture}[remember picture,overlay]
\node[anchor=south,yshift=10pt] at (current page.south) {\fbox{\parbox{\dimexpr0.65\textwidth-\fboxsep-\fboxrule\relax}{\submittedtext}}};
\end{tikzpicture}%
}

\newcommand\copyrighttext{%
  \footnotesize \textcopyright \the\year{} IEEE. Personal use of this material is permitted. Permission from IEEE must be obtained for all other uses, including reprinting/republishing this material for advertising or promotional purposes, collecting new collected works for resale or redistribution to servers or lists, or reuse of any copyrighted component of this work in other works.}

\begin{document}

\title{Scalable Asynchronous Single Flux Quantum Up-Down Counter using Josephson Trapping Lines and $\upalpha$-Cells
}

\author{\IEEEauthorblockN{Mustafa Altay Karamuftuoglu, Beyza Zeynep Ucpinar, Sasan Razmkhah and Massoud Pedram, \textit{IEEE Fellow}}\\

\thanks{
This work is supported by the National Science Foundation (NSF) under the project Expedition: (Design and Integration of Superconducting Computation for Ventures beyond Exascale Realization) with grant number 2124453.
(Corresponding author: M. A. Karamuftuoglu)

M. A. K., B. Z. U., S. R., and M. P. are with the Ming Hsieh Department of Electrical and Computer Engineering, University of Southern California, Los Angeles, USA. (karamuft@usc.edu, ucpinar@usc.edu, razmkhah@usc.edu pedram@usc.edu)
}
}

\maketitle

\submittednotice

\begin{abstract}
We present a scalable, clockless up-down counter architecture implemented using single-flux quantum (SFQ) logic to enable efficient state management in superconductor digital systems. The proposed design eliminates the reliance on clocked storage elements by introducing the Josephson Trapping Line (JTrL). This bidirectional pulse-trapping structure enables persistent, non-volatile state storage without clocking. The counter integrates $\upalpha$-cells with a splitter (SPL) element to make bidirectional data propagation possible and support multi-fanout connectivity. The design supports increment, decrement, and read operations and includes a control unit that guarantees correct output behavior across all valid state transitions. Circuit-level simulations based on SPICE models demonstrate robust bidirectional functionality across a 3-bit state range [-4 to +4] at an operating frequency of 4 GHz. The proposed counter offers a modular and scalable solution suitable for integration into larger superconducting systems targeting quantum computing, neuromorphic processing, and cryogenic sensing applications.
\end{abstract}


\section{Introduction}

Up-down counters are crucial in digital electronics, arithmetic operations, control mechanisms, and event tracking. These counters are widely used in frequency division, digital signal processing (DSP), microprocessor register control, digital filtering, and neural network architectures \cite{stan1998long}. A typical up-down counter increments or decrements its count based on control signals, making it essential for tracking dynamic changes in various digital systems. There are several works on CMOS-based down counters. \cite{sync_updown_cmos} proposes a pre-scaled synchronous up-down counter with a clock period independent of counter size, enabling scalability without speed loss. In \cite{cmos_sar_adc}, the authors present a compact 4-bit successive approximation register (SAR) analog-to-digital converter (ADC) that uses an up-down counter as SAR logic, achieving low power and high speed for health care applications. The study in \cite{cmos_updown_2} shows that a high-speed, low-power up-down counter is designed in 0.18-$\mu$m CMOS, demonstrating strong performance and efficiency for embedded systems.

Furthermore, several other implementations use memristor technology and memristor-CMOS hybrid technology. In \cite{memristor_updown}, the authors introduce a memristor-based up-down counter using material implication logic, achieving high speed, low power, and compact area through non-volatile T flip-flops. \cite{hbrid_memristor_cmos} proposes a 4-bit hybrid memristor-CMOS up-down counter using memristor-ratioed logic, achieving reduced transistor count, lower power, and improved area efficiency. Although CMOS-based up-down counters are standard, they face power inefficiencies and operational speed constraints at high frequencies. Emerging designs using memristor or hybrid memristor-CMOS technologies offer improvements but are challenged by fabrication complexity, integration issues, limited resolution, and technology immaturity.

Superconductor electronics (SCE) \cite{likharev1991a, razmkhahBook} aims to overcome the limitations of conventional semiconductor technology by offering ultra-fast, energy-efficient digital circuits at cryogenic temperatures. These circuits use the unique quantum properties of Josephson junctions (JJs), which allow low-power dissipation, rapid switching times, and nearly lossless signal transmission. Among various superconducting logic families, single-flux quantum (SFQ) logic, including rapid SFQ (RSFQ) and energy-efficient RSFQ (ERSFQ) \cite{energy_efficient_sfq}, has gained significant attention due to its potential to operate at multi-GHz frequencies with extremely low power consumption. These features make SFQ logic a viable candidate for next-generation computing systems for high-speed data processing.

The motivation behind this work arises from the demand for high-speed, energy-efficient digital counters that can be integrated into superconducting computing architectures. SFQ-based circuits offer a compelling alternative to CMOS due to their ultra-fast switching capability and low power dissipation. Thus, by implementing a superconducting up-down counter using SFQ logic, we focus on energy efficiency, high performance, and ease of integration into SFQ-based computing platforms.

Several superconductor-based up-down counter designs are implemented in both SFQ and AQFP technologies. Authors in \cite{sfq_updown_dynamic_range} present a digital double relaxation oscillation SQUID (DROS) system with an 8-bit SFQ up-down counter, offering high sensitivity and wide dynamic range for flux digitization. In \cite{sfq_updown_neuralcomp_stoch}, the authors introduce a high-speed SFQ-based down counter for neural network computation using stochastic logic, achieving a constant accumulation speed independent of bit width. \cite{aqfp_updown} presents binary up-down counters based on AQFP logic, demonstrating ultra-low power consumption. Although binary SFQ counters, frequency dividers, and shift registers are shown in prior research, SFQ-based up-down counter architectures remain relatively unexplored. Most existing designs focus on ripple counters or synchronous binary counters. Still, there is a lack of efficient bidirectional counting mechanisms that can asynchronously handle high-speed pulse-driven operations. Our work addresses this gap by proposing a scalable, energy-efficient up-down counter for superconducting digital systems, ensuring accuracy, low power consumption, and high-speed performance.

Unlike conventional designs that mainly rely on D flip-flops (DFF), toggle flip-flops (TFF), or non-destructive readout memory (NDRO) cells, this work introduces an asynchronous counter. The design improves area efficiency and operational robustness by using the bidirectional behavior of superconducting circuits. The counter incorporates the \textit{Josephson Trapping Line} (JTrL) as a bidirectional pulse storage mechanism, eliminating the need for a clock and enabling persistent pulse storage, supporting trapped-state behavior. Integrating $\upalpha$-cells simplifies the design by facilitating pulse propagation and path sharing, reducing overall complexity. The counter operates with mutually exclusive increment, decrement, and read control signals while managing overflow conditions through $\upalpha$ cells. The correct functionality has been demonstrated across the [+4, -4] state range with simulations at 4 GHz using JSIM \cite{nakamura2009a}.

The key contributions of this paper are as follows.
\begin{itemize}
\item Demonstrating an asynchronous up-down counter with robust operation at 4 GHz;
\item Integrating the $\upalpha$-cell with standard SFQ digital library cells to enable bidirectional pulse propagation;
\item Introducing a clockless pulse trapping mechanism using the JTrL for state machine applications; and
\item Proposing a scalable state circuit architecture, allowing additional JTrLs and $\upalpha$-cells to be incorporated without increasing design complexity.
\end{itemize}

\section{Device Modeling}
\begin{figure} [!htbp]
  \centering
  \includegraphics[width=1\linewidth]{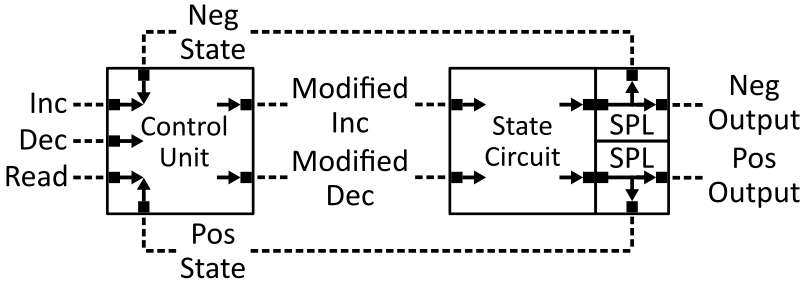}
  \caption{Integration of up-down counter. The up-down counter has three mutually exclusive control signals: Inc for incrementing the state, Dec for decreasing the state, and Read for performing a read operation. Directly applying such signals to the state circuit may introduce erroneous outputs. The up-down counter behavior can be achieved correctly by incorporating the control unit.}
  \label{fig:Integration}
\end{figure}

The up-down counter comprises two main components: the state circuit and the control unit, as shown in Fig. \ref{fig:Integration}. The state circuit defines a finite set of states and controls their transitions based on external inputs and internal logic. In this design, the state circuit implements a Mealy finite state machine (FSM), where the output depends on both the current state and the input signals.

Based on external control signals, the control unit determines whether the counter value increases or decreases. Specifically, it processes the increment (Inc), decrease (Dec), and read (Read) signals to update the counter state accordingly. In addition, it ensures the correct transition through the state sequence and generates the corresponding output pulse sequence.

The design operates without a clock signal, making it inherently asynchronous. While bidirectional DFFs may be used to introduce pipelining and synchronization, this paper focuses solely on the counter's asynchronous operation. Notice that half of the JTLs and intermediate delay elements have been removed from the diagrams to simplify visualization in the following sections.

\subsection{State Circuit}
\begin{figure} [!htbp]
  \centering
  \includegraphics[width=1\linewidth]{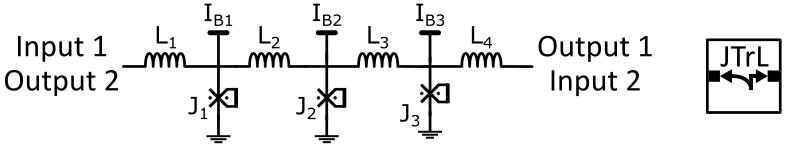}
  \caption{Josephson Trapping Line with unidirectional trap. The first pulse from Input 1 is trapped in the J1–L2–J2 storing SQUID loop as a persistent circulating current, allowing subsequent pulses to propagate. Pulses from Input 2 either neutralize the stored current or propagate to Output 2. The unidirectional trapping behavior can be extended by increasing L3 in the J2–L3–J3 loop to enable bidirectional trapping. Circuit parameter values for the unidirectional configuration are as follows: L1=0.93 pH, L2=7.64 pH, L3=2.12 pH, L4=1.31 pH, J1=303 $\mu$A, J2=318 $\mu$A, J3=282 $\mu$A, IB1=237.64 $\mu$A, IB2=125.25 $\mu$A, IB3=182.35 $\mu$A.}
  \label{fig:JTrL}
\end{figure}

A superconducting state circuit requires a SQUID structure capable of storing pulses. Many conventional designs use DFF, TFF, or NDRO cells that depend on a clock signal for storage. Instead, we introduce the Josephson Trapping Line (JTrL), which supports bidirectional operation without clocking. Structurally, the JTrL follows the Josephson Transmission Line (JTL) design, but one of its SQUIDs is modified for pulse storage by increasing its inductance. The corresponding schematic and its symbol are shown in Fig. \ref{fig:JTrL}.

In Fig. \ref{fig:JTrL}, the first incoming pulse from the trapping side (Input 1) enters the storage SQUID loop (J1–L2–J2) and gets trapped, allowing subsequent pulses to propagate. Notice that the bias current IB2 and the circulating current are insufficient to trigger J2, keeping the pulse stored in the loop. As a result, once the first pulse is trapped, the second pulse propagates freely, while the first pulse remains a persistent circulating current within the storage loop. Conversely, pulses from Input 2 initially propagate through the J3–L3–J2 SQUID loop. If the storage loop contains a circulating current, the first pulse from Input 2 neutralizes this current without producing an output. On the other hand, if the storage loop is empty, the Input 2 pulse propagates directly to Output 2. Increasing L3, similar to L2, allows the J2–L3–J3 SQUID loop to function as a storage loop, enabling bidirectional trapping within the design.

\begin{figure} [!htbp]
  \centering
  \includegraphics[width=1\linewidth]{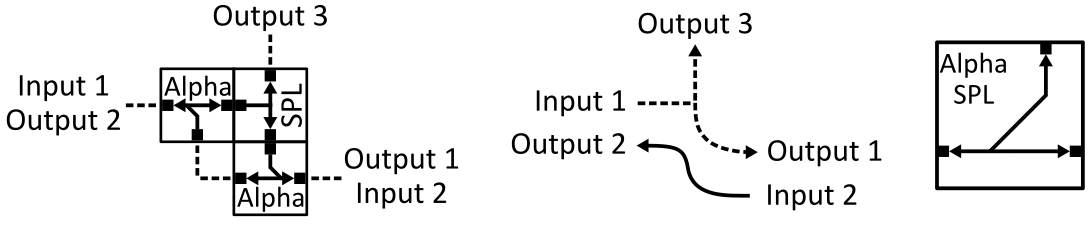}
  \caption{Bidirectional multifanout line with $\upalpha$-SPL.}
  \label{fig:AlphaSPL}
\end{figure}

The $\upalpha$-cell is an interconnect capable of propagating bidirectional data. We introduce a multifanout line called the $\upalpha$-SPL to enable bidirectional operation within the state circuit. The design consists of one SPL and two $\upalpha$-cells, where the $\upalpha$-cell combines the functionalities of a Josephson Transmission Line (JTL) and a Confluence Buffer (CBU) while increasing the fan-in of the bidirectional line. The SPL facilitates multifanout capability, allowing the design to support multiple outputs. The number of SPLs can be adjusted based on application requirements. The bidirectional characteristics of the design enable pulse path sharing, eliminating the need for multiple dedicated datapaths. The corresponding cell view is shown in Fig. \ref{fig:AlphaSPL}. An SFQ pulse at Input 1 generates output pulses at Output 1 and 3. In contrast, an SFQ pulse from Input 2 propagates only to Output 2.

\begin{figure} [!htbp]
  \centering
  \includegraphics[width=1\linewidth]{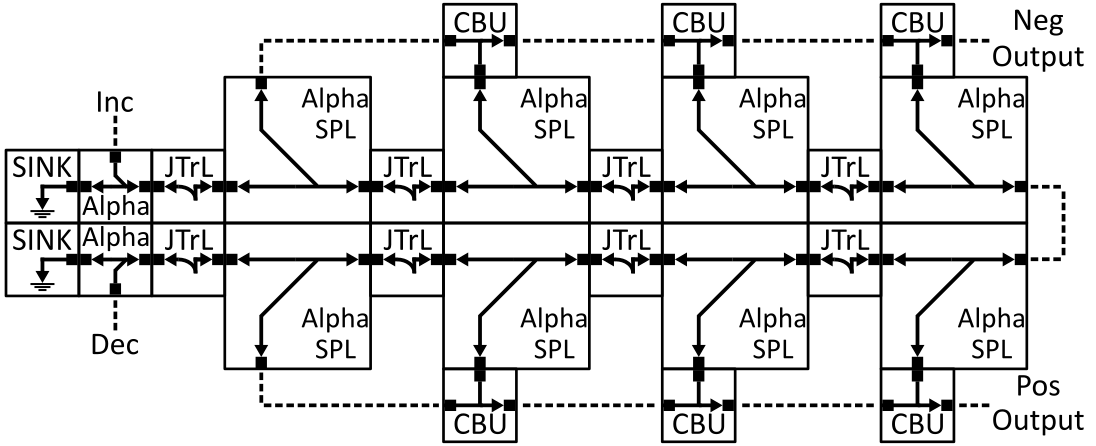}
  \caption{State Circuit. Pulses from Inc neutralize any trapped Dec pulses at the upper half of the design. Inc pulses get trapped in the bottom half instead and generate output through the $\upalpha$-SPL if no Dec pulses are present.}
  \label{fig:StateCircuit}
\end{figure}

To build the state circuit, we combine $\upalpha$-SPLs with JTrL cells to control the storage and removal of pulses. Inc pulses arriving at the circuit eliminate any previously stored Dec pulses. If no Dec pulses are present in the upper JTrLs, the Inc pulses move to the lower JTrLs, where they get trapped. Before trapping occurs in JTrL, $\upalpha$-SPL generates an output where each pulse reflects the number of stored pulses. Our implementation includes four JTrLs for positive states and four JTrLs for negative states, defining a [-4, +4] state range, which fits within a 3-bit representation. The counter range can be extended by incorporating more JTrLs and $\upalpha$-SPLs. The design is shown in Fig. \ref{fig:StateCircuit}.

We integrate $\upalpha$ and sink cells, which are used for grounding the SFQ output with matched impedance, on the input side for the Inc and Dec pins for the upper and lower halves. When all JTrLs are fully occupied, the system reaches its maximum state, and any further pulses can no longer be trapped. In this case, the excess pulse propagates through the $\upalpha$-cell and is directed to the sink cell. The unidirectional input pin of the $\upalpha$-cell handles Inc and Dec pulses without interrupting the overall functionality.

In the design shown in Fig. \ref{fig:StateCircuit}, output generation occurs only when an Inc input is applied while the state is in [0, +4] or when a Dec input is applied while the state is in [-4, 0]. As a result, no output is produced if an Inc input neutralizes a stored pulse in the upper half or if a Dec input removes a pulse from the bottom half. The state circuit requires a control unit that manages pulse generation accordingly to enable output generation in these cases.

\subsection{Control Unit}
\begin{figure} [!htbp]
  \centering
  \includegraphics[width=1\linewidth]{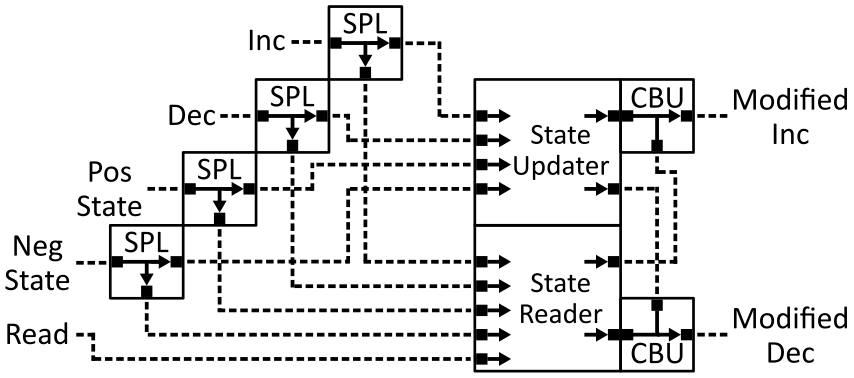}
  \caption{Control Unit.}
  \label{fig:ControlUnit}
\end{figure}

The control unit includes two subcircuits: the state updater and the state reader. The state updater ensures the output generation when Inc is applied in the negative state [-4, -2] or Dec is applied in the positive state [+2, 4], where the output would not normally occur. The state reader handles read operations when neither Inc nor Dec is active. Both subcircuits generate Modified Inc and Modified Dec signals, which are combined by the CBU, as shown in Fig. \ref{fig:ControlUnit}. Output generation in these circuits is mutually exclusive, as the State Updater operates based on Inc and Dec signals, while the State Reader depends on the Read control signal.

\subsubsection{State Reader}
Since the state circuit only operates with Inc and Dec, introducing an additional read signal would increase the complexity of the design. Therefore, simultaneously applying Inc and Dec signals can perform the read operation. If the system is in a positive state: [+1, +4], the Dec signal removes a single pulse stored in a JTrL before the Inc signal arrives. Due to the symmetrical structure of the state circuit, Dec reaches the cell earlier than Inc, resulting in a decrease-then-increase operation. The process applies to negative states, where the sequence increases and decreases.

\begin{figure} [!htbp]
  \centering
  \includegraphics[width=0.70\linewidth]{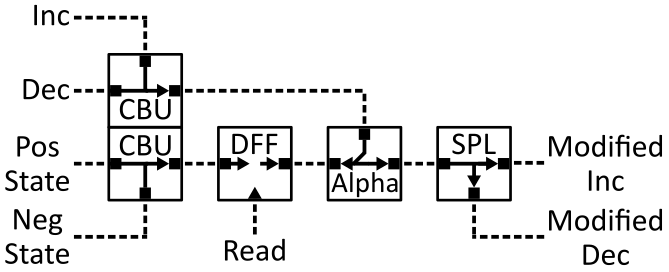}
  \caption{State reader circuit. The circuit reads when the state is nonzero, as Pos-State and Neg-State set the DFF. When the state is zero, no input arrives from the Pos-State or Neg-State, keeping the DFF output at zero. Inc/Dec pins are reserved for the next operation, and any previously stored DFF value is cleared if the next operation is Inc or Dec, regardless of the prior operation.}
  \label{fig:StateReader}
\end{figure}

However, simultaneously applying Inc and Dec signals introduces uncertainty when the state is zero. Since no pulses are trapped in JTrLs, Inc and Dec meet without eliminating a stored pulse. The outcome depends on which pulse triggers a junction first at the intermediate point of the data path, making it highly sensitive to propagation delays. The read signal should not be applied when the state is zero to avoid unpredictable behavior, as no output is expected. Hence, the read signal timing requires a detection mechanism to determine whether the state is nonzero and suitable for a read operation. The corresponding design is shown in Fig. \ref{fig:StateReader}.

The Pos-State and Neg-State signals serve as outputs of the state circuit, representing the current state of the counter. Since these signals are mutually exclusive, we use a CBU cell to perform parallel-to-serial conversion, directing the corresponding output to the DFF input. This setup allows the state signals to set the DFF state.

Applying a Read signal to the DFF clock pin enables the generation of Modified Inc and Modified Dec signals while ensuring no increment or decrement occurs when the state is zero. At the DFF output node, an $\upalpha$-cell is implemented to provide an asynchronous reset for the DFF. The $\upalpha$-cell's unidirectional input combines Inc and Dec signals, clearing the DFF state set in previous cycles.

\subsubsection{State Updater}
\begin{figure} [!htbp]
  \centering
  \includegraphics[width=0.62\linewidth]{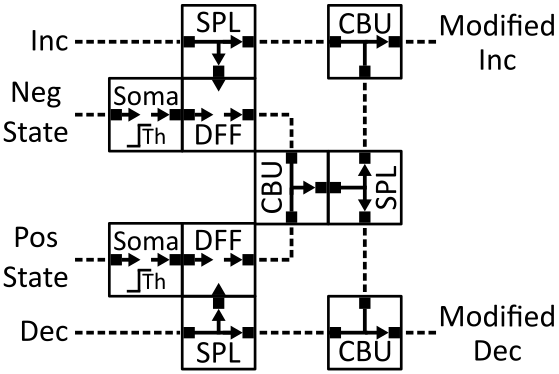}
  \caption{State Updater. Since the up-down module cannot generate an output when the state is negative ($<-1$), and the control input is increment (also vice versa), we increment first and then perform read operation by generating both Modified Inc and Modified Dec signals in addition to the initial inc signal. The single-input Soma performs threshold function checks if the output of the previous operation had an amplitude greater than 1. If the amplitude is 1, the increment or decrement operation will bring the state to 0, and the read operation is unnecessary.}
  \label{fig:StateUpdater}
\end{figure}

The state circuit does not produce an output when Inc is applied in a negative state or Dec is applied in a positive state. The state updater generates Modified Inc and Modified Dec signals to address this limitation, which are provided to the state circuit. Since the read operation is performed by applying Inc and Dec in non-zero states, the state updater ensures that the circuit follows an increment-then-read sequence for negative states and a decrement-then-read sequence for positive states. The cell view of the design is given in Fig.~\ref{fig:StateUpdater}.

Inc and Dec signals propagate through SPL cells to initiate the increment or decrement operation, setting the circuit to the desired state. A Soma cell, described in \cite{karamuftuoglu2023jj}, is paired with a DFF cell to enable the read operation in the cases described in the State reader section. The Soma cell has a threshold of two SFQ pulses when acting as a threshold gate, ensuring that it only generates an output when the state amplitude exceeds one and provides multiple arbitrary SFQ pulses, at which point it sets the DFF state.

In this configuration, the Inc signal acts as the clock for Neg-State signals after the threshold cell, while the Dec signal serves as the clock for Pos-State signals. When a DFF cell is set by the state signals, confirming that the state amplitude is greater than one, the corresponding DFF generates an output upon receiving an Inc or Dec signal. This output pulse is then split by an SPL cell, generating SFQ pulses on both Modified Inc and Modified Dec outputs, thereby executing the read operation.

\section{Simulation Result}
Each major circuit block is simulated using JSIM \cite{nakamura2009a} to verify the proposed design's overall behavior. These simulations confirm the correct functionality of individual components and their integration into a complete asynchronous up-down counter.

\begin{figure} [!htbp]
  \centering
  \includegraphics[width=0.76\linewidth]{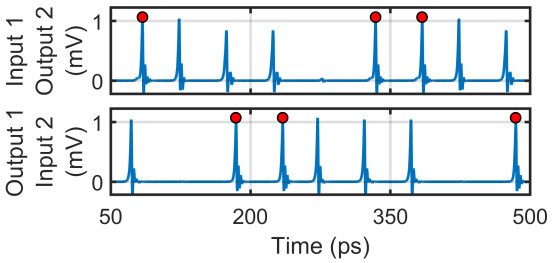}
  \caption{Simulation of the JTrL, showing unidirectional trapping behavior. The first pulse from Input 1 is trapped, and subsequent pulses propagate. A pulse from Input 2 clears the trapped pulse, confirming the correct trap-release operation and persistent storage. The output pulses are marked with red.}
  \label{fig:JTrLSim}
\end{figure}

The first simulation targets the JTrL, which is the state circuit's primary pulse storage element. As shown in Fig. \ref{fig:JTrLSim}, the JTrL operates in a unidirectional trapping mode. A pulse from Input 1 is trapped in the SQUID loop while subsequent pulses propagate through the circuit. Pulses from Input 2 propagate directly unless a pulse from Input 1 is already stored. In such cases, the first Input 2 pulse neutralizes the trapped pulse, and the following Input 2 pulses propagate. In the waveform, the first Input 2 pulse is observed at Output 2, confirming the expected propagation. The first Input 1 pulse does not appear at Output 1, indicating successful trapping, while the second Input 1 pulse appears at approximately 175 ps. The Input 2 pulse at 275 ps does not produce an output, validating the trap-release behavior. The following Input 2 pulse appears at Output 2, confirming that the path has been cleared.

For the JTrL cell, we utilize unidirectional trapping, which means that one pulse from Input 1 will be trapped in the SQUID loop, and the consecutive ones will propagate. However, the pulses coming from Input 2 will propagate directly through the cell. In case of a trapped pulse coming from Input 1, the first Input 2 pulse will remove the trapped one, allowing the consecutive ones to propagate. In the simulation, the first pulse from Input 2 appears at Output 2. However, the first pulse of Input 1 does not appear in Output 1, indicating that it is trapped, whereas the consecutive pulse at 175 ps appears in Output 1.

\begin{figure} [!htbp]
  \centering
  \includegraphics[width=0.76\linewidth]{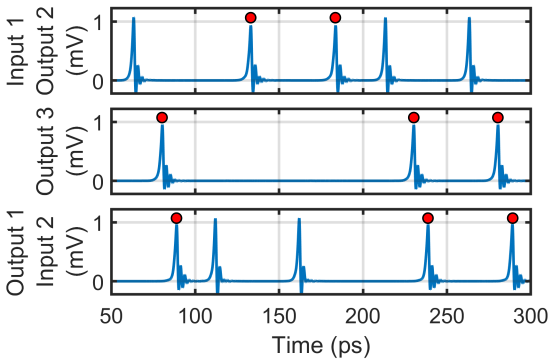}
  \caption{Simulation of the $\upalpha$-SPL cell, demonstrating bidirectional propagation and multifanout behavior. A pulse from Input 1 reaches Outputs 1 and 3, while a pulse from Input 2 propagates only to Output 2, validating directional control and fanout functionality. The output pulses are marked with red.}
  \label{fig:AlphaSPLSim}
\end{figure}

The $\upalpha$-SPL cell is then evaluated to demonstrate bidirectional propagation and multifanout behavior. In Fig. \ref{fig:AlphaSPLSim}, a pulse from Input 1 generates outputs at both Output 1 and Output 3 before 100 ps. A slight delay is observed at Output 1 due to the additional path through the $\upalpha$-cell. Pulses arriving from Input 2 (e.g., at 110 ps and 160 ps) appear only at Output 2, confirming the directional routing based on the input port.

\begin{figure} [!htbp]
  \centering
  \includegraphics[width=0.77\linewidth]{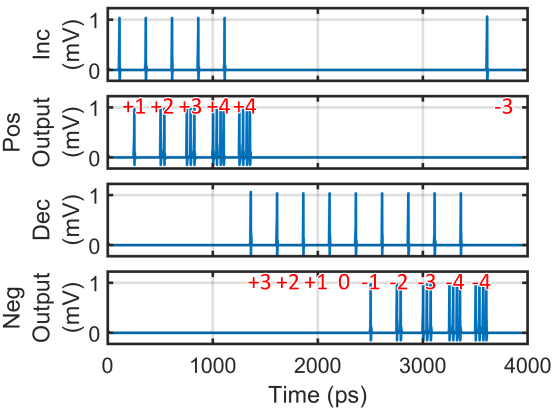}
  \caption{State circuit simulation for a [-4, +4] range, showing correct increment and decrement behavior. Inc pulses are stored in JTrLs and generate output when transitioning from 0 to +4. Dec pulses only generate output after crossing into the negative state, confirming correct pulse-based state tracking.}
  \label{fig:StateCircuitSim}
\end{figure}

The state circuit, composed of JTrLs and $\upalpha$-SPLs, is tested for functional correctness using the design with [-4, +4] range shown in Fig. \ref{fig:StateCircuit}. The waveform in Fig. \ref{fig:StateCircuitSim} shows the circuit incrementing its state using Inc pulses. Each incoming pulse is stored in a distinct JTrL until the +4 limit is reached, producing output pulses during this phase. While in the positive state, the Dec input pulse will not generate any output. Once the state reaches zero, output pulses appear with each Dec pulse, confirming proper decrement and state tracking.

\begin{figure} [!htbp]
  \centering
  \includegraphics[width=0.77\linewidth]{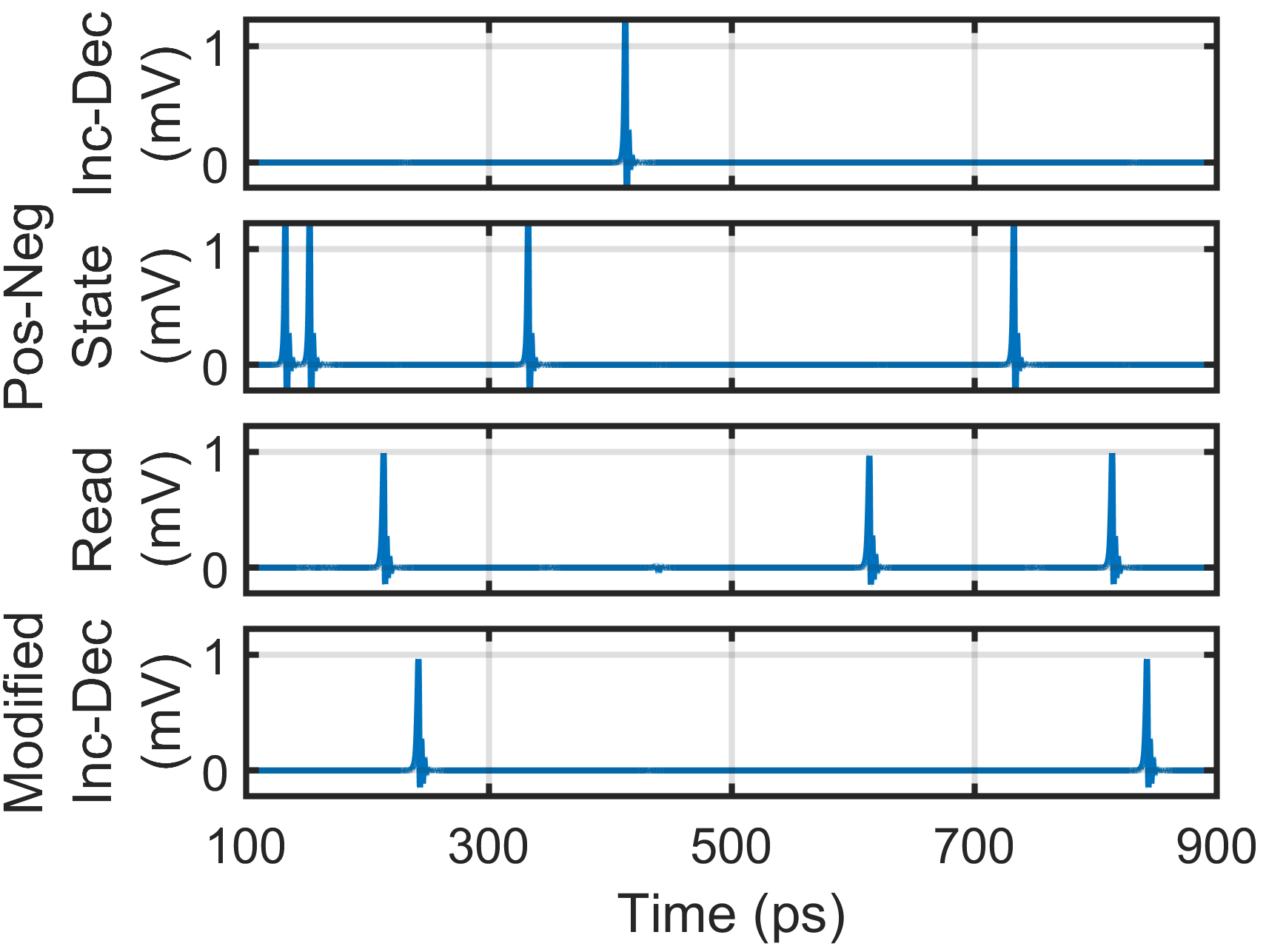}
  \caption{Simulation of the state reader circuit. The circuit correctly suppresses output at zero state and generates output only when a read signal aligns with a valid stored pulse in DFF.}
  \label{fig:StateReaderSim}
\end{figure}

The state reader circuit supports the read operation and is simulated with JSIM. This circuit uses a DFF to capture the last state, triggered by a Read signal acting as the clock. The Pos-State and Neg-State signals indicate that the system is in a nonzero state. In Fig. \ref{fig:StateReaderSim}, the combined CBU output from these signals is labeled Pos-Neg State, while the CBU output from Inc/Dec is labeled Inc-Dec. The DFF output before the SPL cell is referred to as Modified Inc-Dec.

The simulation begins with two pulses from the Pos-Neg State, followed by a Read signal at around 200 ps, triggering the DFF. This process results in a pulse at Modified Inc-Dec, which then applies signals to both Inc and Dec inputs of the state circuit. Between 300 and 500 ps, there is no Read signal, and no output is generated. After 500 ps, with Pos-Neg State inactive (indicating a zero state), applying Read results in no output, as expected. A valid read operation occurs again after 700 ps when a single-state pulse is present.

\begin{figure} [!htbp]
  \centering
  \includegraphics[width=0.77\linewidth]{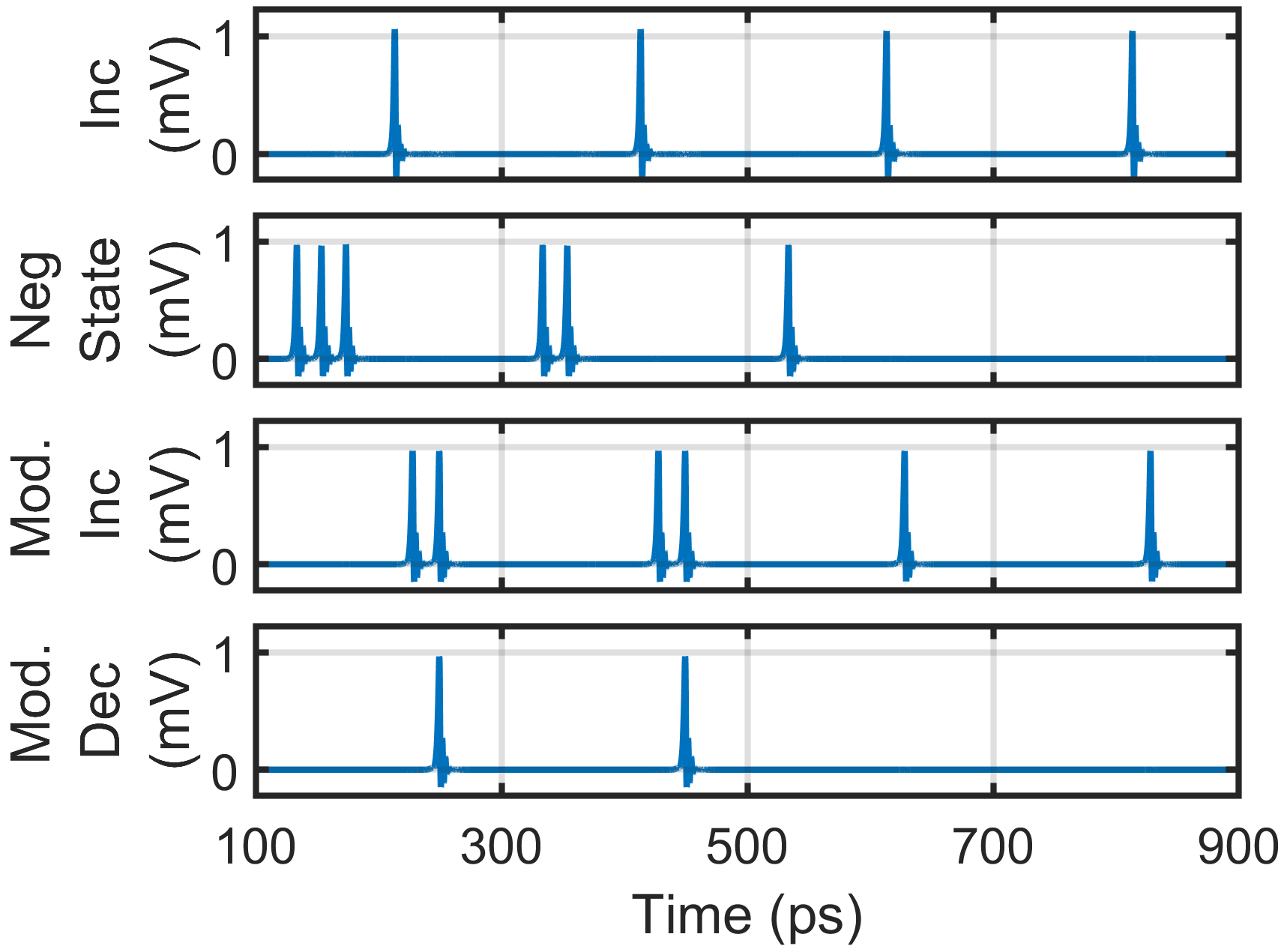}
  \caption{Simulation of the state updater circuit. The circuit handles read-after-increment for negative states and correctly halts read behavior at state 0.}
  \label{fig:StateUpdaterSim}
\end{figure}

We require the state updater circuit to generate output in cases where the state circuit alone does not respond, such as Inc in a negative state or Dec in a positive state. The corresponding simulation in Fig. \ref{fig:StateUpdaterSim} focuses on negative states with Inc applied. The initial state is set to -3, and an Inc pulse results in two pulses on Modified Inc and one on Modified Dec, representing an increment followed by a read operation. The same behavior is observed at state '-2'. After 500 ps, this behavior stops once the state reaches 0 and then +1, as no read is required in the nonnegative region.

\begin{figure} [!htbp]
  \centering
  \includegraphics[width=1\linewidth]{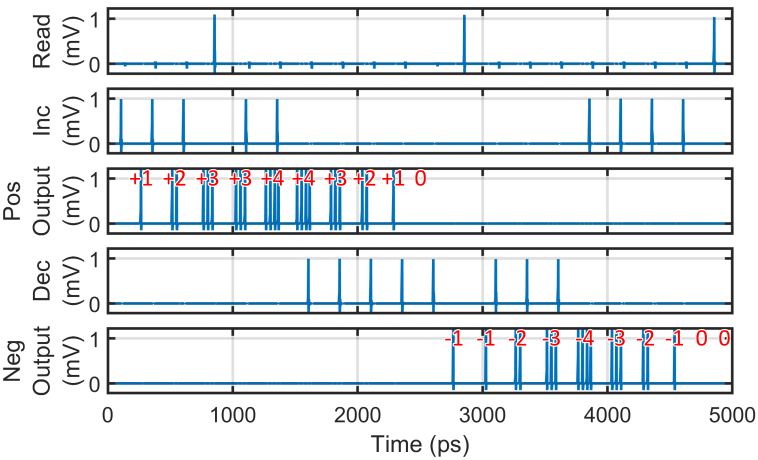}
  \caption{Full design simulation of the asynchronous up-down counter across the complete [-4, +4] range. The counter performs accurate state updates and output generation under increment, decrement, and read signals at 4 GHz, confirming the functional correctness of the complete architecture.}
  \label{fig:UpDownCounterSim}
\end{figure}

Finally, the complete asynchronous up-down counter is simulated by combining all blocks. The waveform in Fig. \ref{fig:UpDownCounterSim} shows a correct behavior in the [-4, +4] range. The simulation begins at state 0, incrementing to +4 with observed outputs during the read phase. Dec pulses are then applied, bringing the state down to -4, followed by Inc pulses to return the counter to 0. The circuit behavior is validated at 4 GHz, confirming the functional correctness of the full design.

\section{Applications and Discussions}
The proposed asynchronous up-down counter supports a wide range of applications in superconducting digital systems, including state tracking, event counting, and event-driven operations. The design eliminates the area and complexity overhead typically introduced by clock tree implementation by operating without a clock signal. As a result, the absence of clocks improves the layout's efficiency and power consumption. The counter's inherent characteristics enable reliable performance in high-speed and low-power cryogenic environments.

Asynchronous counters offer valuable functionality in quantum computing systems \cite{semenovSFQcontrolQubit}. SFQ pulses are used in the architectures to implement fast and energy-efficient logic operations. Up-down counters provide essential support during qubit control and measurement. Such counters enable precise modulation of quantum states without relying on continuous analog feedback by using SFQ-based logic to count and track discrete events within a qubit control cycle.

Neuromorphic computing also benefits from asynchronous counters. Spiking neural networks based on superconducting logic use counters as spike accumulators or state monitors \cite{stdpKaramuft}. Up-down counters track excitatory and inhibitory spikes, supporting timing and learning rules in these systems. This functionality supports the development of large-scale energy-efficient neural networks that mimic the parallel and event-driven behavior of the human brain. The counter’s asynchronous nature aligns well with timing-based neural dynamics, making it a strong candidate for scalable cryogenic neuromorphic architectures.

Asynchronous up-down counters also play a critical role in data acquisition and superconducting sensor systems. Counters generate sequential addresses that synchronize with sensor outputs in the context of detector arrays. This approach significantly reduces wiring complexity and improves scalability, making it well-suited for high-resolution imaging systems \cite{sfq_address_encoder}. Counters are equally crucial in digital SQUID readout architectures, where they function as digital integrators. By counting discrete flux quanta instead of relying on analog loops, up-down counters enable real-time reconstruction of analog signals with high resolution \cite{digital_squid}. These examples highlight how asynchronous counters support precise, event-based operation across superconducting applications, particularly where compact design, energy efficiency, and timing accuracy are critical.

\section{Conclusion}
This work presents a novel asynchronous up-down counter that eliminates the need for clocked storage cells while leveraging the bidirectional behavior of superconducting circuits to improve area efficiency and robustness. The design introduces the Josephson trapping line as a bidirectional pulse storage mechanism, enabling persistent state retention without a clock. Furthermore, integrating $\upalpha$-cells facilitates pulse propagation and path sharing, reducing overall complexity. The scalability of the design allows the counter range to be extended by adding JTrLs and $\upalpha$-cells without increasing the design complexity. The simulation results at 4 GHz using JSIM confirm the correct operation of the counter within a [-4, +4] state range. The techniques introduced provide a foundation for scalable, high-speed superconducting state machines with bidirectional operations, contributing to the advancement of SFQ-based digital systems.

\bibliographystyle{IEEEtran}
\bibliography{references}
\end{document}